\documentclass{IEEEtran}
\usepackage[T1]{fontenc}
\usepackage[latin9]{inputenc}
\usepackage{amsthm}
\usepackage{amsbsy}
\usepackage{amstext}
\usepackage{amssymb}
\usepackage{graphicx}
\usepackage{esint}
\usepackage{subeqn}

\makeatletter

\providecommand{\tabularnewline}{\\}

\theoremstyle{plain}
\newtheorem{theorem}{Theorem}
\theoremstyle{remark}
\newtheorem{remark}{Remark}
\theoremstyle{plain}
\newtheorem{proposition}{Proposition}

\@ifundefined{showcaptionsetup}{}{%
 \PassOptionsToPackage{caption=false}{subfig}}
\usepackage{subfig}
\makeatother

\begin{document}

\title{Two Design Issues in Cognitive Sub-Small Cell for Sojourners}

\author{Xin Jin, Abdelwaheb Marzouki, Djamal Zeghlache, \textsl{Member},
\textsl{IEEE}, Linghe Kong, \textsl{Member}, \textsl{IEEE}, and Athanasios
V. Vasilakos, \textit{Senior Member}, \textsl{IEEE }%
\thanks{X. Jin is with Pierre-and-Marie-Curie University (University Paris
6) and Institut Mines-Telecom, Telecom SudParis, CNRS Samovar UMR
5157, France. (e-mail: felixxinjin@gmail.com). A. Marzouki and D.
Zeghlache are with Institut Mines-Telecom, Telecom SudParis, CNRS
Samovar UMR 5157, France. 

L. Kong is with Shanghai Jiao Tong University, China and Singapore
University of Technology and Design, Singapore.

A. V. Vasilakos is with the Dept. of Computer and Telecommunications
Engineering, University of Western Macedonia, Greece.%
}}
\maketitle
\begin{abstract}
In this paper, we propound a solution named Cognitive Sub-Small Cell
for Sojourners (CSCS) in allusion to a broadly representative small
cell scenario, where users can be categorized into two groups: sojourners
and inhabitants. CSCS contributes to save energy, enhance the number
of concurrently supportable users and enshield inhabitants. We consider
two design issues in CSCS: i) determining the number of transmit antennas
on sub-small cell APs; ii) controlling downlink inter-sub-small cell
interference caused by uncertain CSI. For issue i), we excogitate
an algorithm helped by the probability distribution of the number
of concurrent sojourners. For issue ii), we propose an interference
control scheme named \textsl{BDBF}: Block Diagonalization (BD) Precoding
based on uncertain channel state information in conjunction with auxiliary
optimal Beamformer (BF). In the simulation, we delve into the issue:
how the factors impact the number of transmit antennas on sub-small
cell APs. Moreover, we verify a significant conclusion: Using \textsl{BDBF}
gains more capacity than using optimal BF alone within a bearably
large radius of uncertainty region.\end{abstract}

\section{Introduction}

Small cells such as femto, pico, and microcells have been aroused
general interest lately due to their consequential effect on enhancing
network capacity, stretching service coverage and cutting back network
energy consumption \cite{Green small-cell,Sleep mode}. 

In many residential and enterprise small cells, according to the duration
of stay, users can be grouped into two categories: sojourners and
inhabitants. Besides the duration of stay, other features can be pinpointed:
the status in the network, the Quality of Service (QoS) requirements,
the geographical location, the regularity or predictability of the
entry and exit time. 

\textbf{Motivated example:}\textbf{\textit{ }}A concrete instance
in real life can well elaborate the above concept. An office area
is covered by a small cell. The staffs and visitors can be classified
as inhabitants and sojourners, respectively. Staffs have fixed entry
and exit time, while visitors enter and depart randomly. Staffs have
higher status, settled and specific QoS requirements, while visitors
do not. The staffs and visitors have their own area to stay in, such
as the working area for staffs and the reception area for visitors.
They also have the common area to stay like the meeting room. 

\textbf{Existing approaches and their limitations:}\textbf{\textit{
}}Multiple-Input Multiple-Output (MIMO) technologies are essential
components in 3GPP Long Term Evolution (LTE)-Advanced \cite{MIMO,MIMO EURASIP}.
Among them, multiuser MIMO technology is markedly advantageous to
enhance the cell capacity especially in highly spatial correlated
channels \cite{generalized channel inversion,R.Chen Multimode,MU MIMO},
since multiuser MIMO technology can serve multiple users simultaneously
on the same frequency. Dirty paper coding (DPC) \cite{Dirty paper,robust dirty paper }
and Block Diagonalization (BD) \cite{BD} are two widely adopted techniques
for multiuser MIMO. DPC is contrived to suit the capacity optimal
requirement \cite{Dirty paper 3}, however, it is hard to implement
owing to its complexity. BD is a practical approach which adopts precoding
to completely eliminate inter-user interference. The deficiency of
this technique is that BD precoding and full rank transmission requires
that the number of transmit antennas is not less than the total number
of receive antennas. Therefore, the number of concurrently supportable
users is restricted by the number of transmit antennas.

\textbf{Cognitive Sub-Small Cell for Sojourners (CSCS) solution:}
The above example and the limitations of existing techniques inspire
us to put forward a solution named CSCS in allusion to the following
small cell scenario. i) Users can be categorized into two groups:
sojourners and inhabitants. ii) Inhabitants have fixed entry and exit
time, while visitors enter and depart randomly. iii) Inhabitants have
higher status, settled and specific QoS requirements. iv) The small
cell coverage area is relatively large. v) The number of users is
relatively large. vi) The overlapping area between the inhabitants'
activity area and the sojourners' activity area is relatively small.
CSCS divides one small cell into two customized sub-small cells for
inhabitants and sojourners respectively for the following reasons.

1) CSCS is a design in line with the trend of green communications
\cite{Green small-cell,green}, which plays a significant role of
saving energy:
\begin{itemize}
\item Each of two sub-small cells is served by its own sub-small cell Access
Point (AP). ISSC serves the inhabitants by Inhabitant Access Point
(IAP). SSSC serves the sojourners by Sojourner Access Point (SAP).
In the case of a small overlap between the two groups' activity areas,
utilizing two sub-small cell APs instead of one single small cell
AP shortens the average distance between User Equipment (UE) and AP.
Accordingly, it reduces the total transmit power.
\item Considering that inhabitants and sojourners may have two different
active timetables, CSCS can turn the sub-small cell AP off outside
the active period of the group which it serves. However, AP of a collective
small cell can only turn itself off when both groups are inactive.
\end{itemize}

2) CSCS enhances the number of concurrently supportable users. We
consider this in the context of BD precoding and full rank transmission.
The number of transmit antennas on AP restricts the number of concurrently
supportable users. The number of transmit antennas on a single AP
is limited by many factors, such as the processing speed of AP and
the size of AP. By utilizing two different sub-small cell APs, CSCS
enables more transmit antennas to be installed in the whole small
cell, thereby enhancing the number of concurrently supportable users.

3) CSCS enshields inhabitants according to the priority. The QoS requirements
and higher status for the inhabitants are ensured by a primary sub-small
cell. The sojourners are served by a secondary sub-small cell. The
primary sub-small cell and the secondary sub-small cell utilize the
same spectrum on condition that the level of interference caused by
the secondary sub-small cell to the primary sub-small cell is kept
tolerable \cite{novel overlay underlay}. 

\textbf{Design challenges:}

\textbf{\textit{1) Determining the number of transmit antennas:}}\textit{
}BD precoding and full rank transmission requires that the number
of transmit antennas is not less than the total number of receive
antennas. Therefore, the number of concurrently supportable users
is limited by the number of transmit antennas. When the number of
users is greater than the number of concurrently supportable users,
AP will select users to simultaneously serve using some kind of user
selection algorithm. 

QoS requirements and the number of users who simultaneously present
make demands on the number of concurrently supportable users. On the
other hand, configuring overmany transmit antennas results in a waste
of resources.

\textbf{\textit{2) Inter-sub-small cell interference:}}\textit{ }Due
to BD precoding, ISSC and SSSC will not cause downlink inter-sub-small
cell interference when perfect IAP-to-sojourner CSI and SAP-to-inhabitant
CSI are acquired by IAP and SAP respectively.

In practice, the perfect acquisition of SAP-to-inhabitant CSI is hampered
by the lack of explicit coordination and full cooperation between
ISSC and SSSC, since inhabitants are scant of willingness to feedback
SAP-to-inhabitant CSI by using their own bandwidth and power. SAP
has to turn to blind CSI estimate or other inexact CSI estimates which
will furnish imperfect SAP-to-inhabitant CSI \cite{A survey of dynamic spectrum access,Yu Zhang Distributed Optimal Beamformers for Cognitive Radios Robust to Channel Uncertainties}.
Therefore, the root cause of interference inflicted by SSSC on ISSC
is uncertain SAP-to-inhabitant CSI. 

In this paper, the interference inflicted by ISSC on SSSC is not a
key focus of our study. We make the assumption that either sojourners
are willing to feedback IAP-to-sojourner CSI by using their own bandwidth
and power to avoid interference owing to perfect IAP-to-sojourner
CSI, or sojourners will not access the spectrum which is heavily used
by the inhabitants. 

\textbf{Main contributions:}

\textbf{\textit{1) Algorithm for determining the number of transmit
antennas:}}\textit{ }We propose an algorithm for determining the number
of transmit antennas on sub-small cell APs helped by the probability
distribution of the number of concurrent sojourners. 

\textbf{\textit{2) BDBF:}}\textit{ }We proposed an interference control
scheme named \textsl{BDBF} for secondary systems which are hampered
from perfect secondary-to-primary Channel State Information (CSI).
We prove and verify that \textsl{BDBF} can gain more capacity than
the interference controller using optimal Beamformer (BF) alone within
a bearably large radius of uncertainty region.

\textbf{\textit{3) Performance Analysis:}} i) From simulation and
numerical results, we find how the factors, such as the standard deviation
of the duration of stay, the mean of the duration of stay, the distribution
of arrival rate, and the total number of sojourners during the observation
time interval, influence the number of concurrent sojourners and further
impact the number of transmit antennas on sub-small cell APs. ii)
We show the benefit of \textit{BDBF}.

The rest of the paper is organized as follows. We present the system
model in Section II. Section III introduces the algorithm for determining
the number of transmit antennas on sub-small cell APs. Section IV
introduces the \textit{BDBF} scheme. We present the simulation results
and provide insights on them in Section V. Finally, we conclude the
paper in Section VI.

\begin{table}
\caption{Notations\label{tab:Notations-1}}
\begin{tabular}{|c|l|}
\hline 
{\scriptsize Symbol} & {\scriptsize Definition}\tabularnewline
\hline 
{\scriptsize $\ulcorner\bullet\urcorner$} & {\scriptsize the ceiling operator}\tabularnewline
{\scriptsize $N_{T}$} & {\scriptsize the number of transmit antennas on the AP}\tabularnewline
{\scriptsize $N_{R}$} & {\scriptsize the number of receive antennas at each UE}\tabularnewline
{\scriptsize $N_{1}$} & {\scriptsize the total number of inhabitants}\tabularnewline
{\scriptsize $N_{2}\left(t\right)$} & {\scriptsize the total number of sojourners at time $t$}\tabularnewline
{\scriptsize $Q$ } & {\scriptsize the minimum QoS requirements of sojourners}\tabularnewline
\textit{\scriptsize $Q'$}{\scriptsize{} } & {\scriptsize the QoS requirements of}\textit{\scriptsize{} }{\scriptsize inhabitants}\tabularnewline
{\scriptsize $N_{U}$} & {\scriptsize the number of concurrently supportable sojourners}\tabularnewline
{\scriptsize $N_{ST}$} & {\scriptsize the possible number of transmit antennas on SAP}\tabularnewline
{\scriptsize $N_{ST}^{*}$} & {\scriptsize the selected number of transmit antennas on SAP}\tabularnewline
{\scriptsize $N_{IT}$} & {\scriptsize the number of transmit antennas on IAP}\tabularnewline
{\tiny $U_{I}=\left\{ I_{1},\ldots,I_{N_{1}}\right\} $} & {\scriptsize the set of inhabitants }\tabularnewline
{\tiny $U_{S(t)}=\left\{ S_{1},\ldots,S_{N_{2}(t)}\right\} $} & {\scriptsize the set of sojourners }\tabularnewline
{\scriptsize $U_{S}$} & {\scriptsize the set of sojourners}\tabularnewline
{\scriptsize $\mathbf{I}_{N}$} & {\scriptsize the identity matrix of size $N\times N$ }\tabularnewline
{\scriptsize $\mathrm{Tr}\left(\bullet\right)$} & {\scriptsize the trace operator}\tabularnewline
{\scriptsize $\left(\bullet\right)^{\mathcal{H}}$} & {\scriptsize Hermitian transpose}\tabularnewline
{\scriptsize $\mathrm{vec\left(\mathbf{X}\right)}$} & {\scriptsize the vector is made up of the columns of matrix $\mathbf{X}$}\tabularnewline
{\scriptsize $\otimes$} & {\scriptsize the Kronecker product}\tabularnewline
\hline 
\end{tabular}
\end{table}

\section{System Model}

We consider downlink transmission in a small cell served by a single
AP. The AP is equipped with $N_{T}$ transmit antennas and every UE
is equipped with $N_{R}$ receive antennas (We listed some notations
used in the paper in Table \ref{tab:Notations-1}.). The AP adopts
$N_{C}$-subcarrier Orthogonal Frequency Division Multiplexing (OFDM),
BD precoding and full rank transmission. Based on BD precoding and
full rank transmission, the number of concurrently supportable users
is determined by $\ulcorner N_{T}/N_{R}\urcorner$ \cite{BD,Low complexity user selection}.
When the number of users in the small cell is greater than the number
of concurrently supportable users, some kind of user selection algorithm
\cite{Low complexity user selection,MU MIMO scheduling} is implemented.

\subsection{Cognitive Sub-Small Cell for Sojourners}

CSCS turns the single small cell into two sub-small cells: Inhabitant
Sub-Small Cell (ISSC) and Sojourner Sub-Small Cell (SSSC), each of
which is served by its own AP. ISSC serves the inhabitants by Inhabitant
Access Point (IAP). SSSC serves the sojourners by Sojourner Access
Point (SAP). Both IAP and SAP exploit BD precoding and full rank transmission.
The sub-small cells use the same number of subcarriers and the same
spectrum as that are used by the small cell AP. We assume that both
sub-small cell APs take inhabitants and sojourners in their respective
coverage area into account to design BD precoding. 

We assume for simplicity that
\begin{itemize}
\item $N_{1}$ inhabitants always stay in the small cell.
\item $N_{2}\left(t\right)$ sojourners stay in the small cell at time $t$.
\item Sojourners arrive the small cell according to a time-varing Poisson
process with arrival rate $\lambda\left(t\right)$ and $N_{2}\left(t=0\right)=0$.
\item The inhabitants' activity area $A_{1}$ is a disk of radius $R_{1}$.
The sojourners' activity area $A_{2}$ is a disk of radius $R_{2}$.
The small cell coverage area $A_{3}$ is a disk of radius $R_{3}$.
$A_{1}$ and $A_{2}$ partially overlap. $A_{3}$ accommodate $A_{1}$
and $A_{2}$ with $R_{3}=\frac{R_{1}\cos w_{1}+R_{2}\cos w_{2}+R_{1}+R_{2}}{2}$,
where $w_{1}$, $w_{2}$ and the positional relationship of $A_{1}$,
$A_{2}$, and $A_{3}$ are illustrated in Figure $1$.
\begin{figure}
\includegraphics[width=8.2cm,height=8.2cm]{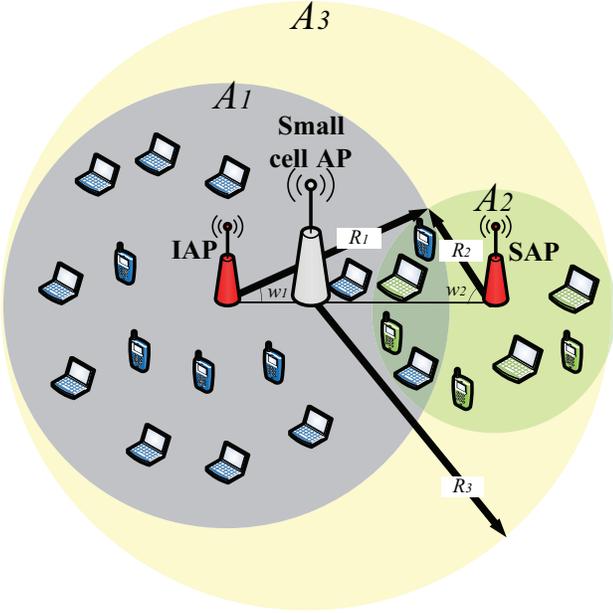}\caption{The system model for CSCS}
\end{figure}

\item The area of $A_{1}$, $A_{2}$, and $A_{3}$ are, respectively, $S_{A_{1}}$,
$S_{A_{2}}$, and $S_{A_{3}}$. The area of overlap between $A_{1}$
and $A_{2}$ is $S_{A_{1}\cap A_{2}}$.
\item $S_{A_{3}}$ is relatively large. $S_{A_{1}\cap A_{2}}$ is relatively
small.
\item Inhabitants and sojourners are uniformly distributed in their respective
activity areas. The number of users is relatively large.
\item The inhabitants have higher status.
\item ISSC and SSSC are respectively consistent with the inhabitants' activity
area $A_{1}$ and the sojourners' activity area $A_{2}$. 
\end{itemize}

\subsection{Cognitive Sub-Small Cell for Sojourners}

CSCS turns the single small cell into two sub-small cells: Inhabitant
Sub-Small Cell (ISSC) and Sojourner Sub-Small Cell (SSSC), each of
which is served by its own AP. ISSC serves the inhabitants by Inhabitant
Access Point (IAP). SSSC serves the sojourners by Sojourner Access
Point (SAP). Both IAP and SAP exploit BD precoding and full rank transmission.
The sub-small cells use the same number of subcarriers and the same
spectrum as that are used by the small cell AP. We assume that both
sub-small cell APs take inhabitants and sojourners in their respective
coverage area into account to design BD precoding.

\subsection{Number of Concurrent Sojourners}

Our analysis for SSSC is based on the following assumptions:
\begin{itemize}
\item Sojourners arrive SSSC according to a time-varing Poisson process
with arrival rate $\lambda\left(t\right)$ and $N_{2}\left(t=0\right)=0$.
\item The duration of stay of one sojourner is modeled as a duration of
a continuous time Markov process with $m+1$ states from its starting
until its ending in the absorbing state, where $m\geq1$. The probability
of this process starting in the state $i$ is $\alpha_{i}$, where
$i=1,\ldots,m+1$. The state $m+1$ is the absorbing state and the
others are transient states. We assume that this process will never
start in the absorbing state, i.e. $\alpha_{m+1}=0$. 
\end{itemize}

The duration of stay of one sojourner has a phase-type distribution
\cite{A new class of models }. We denote by $\mathbf{R}$ the matrix
which contains the transition rates among the transient states. $\mathbf{R}$
is a matrix of size $m\times m$. The duration of stay of one sojourner
is denoted by $X$. The Cumulative Distribution Function (CDF) of
$X$ is given by

\begin{equation}
F_{X}\left(x\right)=1-\boldsymbol{\alpha}\exp\left(\mathbf{R}x\right)\mathbf{1},\; x\geq0,\label{eq:service time distribution}
\end{equation}
where $\boldsymbol{\alpha}=\left[\alpha_{1},\ldots,\alpha_{m}\right]$,
$\exp\left(\bullet\right)$ is the matrix exponential and $\mathbf{1}$
is an $m\times1$ vector with all elements equal to $\text{1}$.
\begin{remark}
Under the above assumptions, SSSC can be modeled as an $M_{t}/G/\infty$
queue with service time follows a phase type distribution, arrival
rate $\lambda\left(t\right)$ and $N_{2}\left(t=0\right)=0$. Authors
in \cite{A. Ghosh Modeling } have given the probability distribution
of the number of concurrent users in such queueing system. Accordingly,
$N_{2}\left(t\right)$ has a Poisson distribution with parameter 
\begin{equation}
v\left(t\right)=\int_{0}^{t}\lambda\left(\tau\right)\left[1-F_{X}\left(t-\tau\right)\right]d\tau,\; t\geq0.\label{eq:users distribution}
\end{equation}
The Probability Mass Function (PMF) of $N_{2}\left(t\right)$ is

\begin{equation}
f_{N_{2}\left(t\right)}\left(n\right)=\mathbb{P}\left(N_{2}\left(t\right)=n\right)=\frac{\left[v\left(t\right)\right]^{n}}{n!}\exp\left(-v\left(t\right)\right),\label{eq:PMF}
\end{equation}
where $n\in\mathbb{Z}_{0}^{+}$. 

The CDF of $N_{2}\left(t\right)$ is

\begin{equation}
F_{N_{2}\left(t\right)}\left(n'\right)=\sum_{n=0}^{n'}f_{N_{2}\left(t\right)}\left(n\right),\label{eq:cumulative distribution function}
\end{equation}
where $n'\in\mathbb{Z}_{0}^{+}$.
\end{remark}

\subsection{Downlink inter-sub-small cell interference}

Due to BD precoding, ISSC and SSSC will not cause downlink inter-sub-small
cell interference when perfect IAP-to-sojourner CSI and SAP-to-inhabitant
CSI are acquired by IAP and SAP respectively.
\begin{remark}
In practice, the perfect acquisition of SAP-to-inhabitant CSI is hampered
by the lack of explicit coordination and full cooperation between
ISSC and SSSC, since inhabitants are scant of willingness to feedback
SAP-to-inhabitant CSI by using their own bandwidth and power. SAP
has to turn to blind CSI estimate or other inexact CSI estimates which
will furnish imperfect SAP-to-inhabitant CSI \cite{A survey of dynamic spectrum access,Yu Zhang Distributed Optimal Beamformers for Cognitive Radios Robust to Channel Uncertainties}.
Therefore, the root cause of interference inflicted by SSSC on ISSC
is uncertain SAP-to-inhabitant CSI. 

In this paper, the interference inflicted by ISSC on SSSC is not a
key focus of our study. We make the assumption that either sojourners
are willing to feedback IAP-to-sojourner CSI by using their own bandwidth
and power to avoid interference owing to perfect IAP-to-sojourner
CSI, or sojourners will not access the spectrum which is heavily used
by the inhabitants. 

Since the IAP-to-inhabitant CSI and SAP-to-sojourner CSI are perspicuous
for IAP and SAP, there is no interference inside ISSC or SSSC in our
system. In some relevant papers about underlay multicarrier cognitive
radio systems with uncertain CSI, the subcarrier scheduling is the
way for steering clear of the interference inside the secondary systems
\cite{T. Al-Khasib Dynamic,Random Subcarrier Allocation}. 
\end{remark}

\section{Determining Number of Transmit Antennas on Sub-small Cell Access
Points}

BD precoding and full rank transmission requires that the number of
transmit antennas is not less than the total number of receive antennas.
Therefore, the number of concurrently supportable users is limited
by the number of transmit antennas. When the number of users is greater
than the number of concurrently supportable users, AP will select
users to simultaneously serve using some kind of user selection algorithm. 

QoS requirements and the number of users who simultaneously present
make demands on the number of concurrently supportable users.

In our system, we express the required number of concurrently supportable
sojourners as $N_{2}\left(t\right)Q$, where $Q$ indicates the minimum
QoS requirements of sojourners, $Q\leq1$, and $N_{2}\left(t\right)Q\in\mathbb{Z}^{+}$.

In this paper, we investigate the issue of determining the number
of transmit antennas under the premise that AP can support the installation
of selected number of transmit antennas.

We put forward a criterion to determine an appropriate number of transmit
antennas on Sub-small Cell Access Points:
\begin{itemize}
\item It pursues the minimum number of transmit antennas which ensures BD
and full rank transmission to be validated.
\item It ensures that the mean value of the probability of the number of
concurrently supportable sojourners being adequate during an applicable
period is no less than the predetermined probability or the growth
of this mean value of probability led by adding one more supportable
sojourner is less than a threshold.
\end{itemize}

We denote by $N_{U}$ the number of concurrently supportable sojourners.
For our system, the probability of the number of concurrently supportable
sojourners being adequate at time $t$ is

\begin{equation}
P\left(N_{U},\; t\right)=F_{N_{2}\left(t\right)}\left(N_{U}/Q\right).\label{eq:probability}
\end{equation}

We define the mean value of the probability of the number of concurrently
supportable sojourners being adequate during the time interval $\left[0,\; T\right]$
as

\begin{equation}
E\left[P\left(N_{U},\; t\right)\right]=\frac{1}{T}\int_{t=0}^{T}P\left(N_{U},\; t\right)dt.\label{eq:mean value of probability}
\end{equation}

According to the first item of our criterion, we have 
\begin{equation}
N_{U}=N_{ST}/N_{R}-N_{1}^{'},\label{eq:Nu NST}
\end{equation}
where $N_{ST}$ is the possible number of transmit antennas on SAP
and $N_{U}\in\mathbb{Z}^{+}$. 
\begin{proposition}
The number of transmit antennas on sub-small cell APs can be determined
as follows:
\end{proposition}
\textit{Since $N_{R}$ and $N_{1}^{'}$ are supposed to be known in
our system model, the selected number of transmit antennas on SAP
$N_{ST}^{*}$ is determined by (\ref{eq:determine N2-1}) or (\ref{eq:determine N2-2}):}

\textit{
\begin{equation}
N_{ST}^{*}=\min_{N_{ST}}\left\{ E\left[P\left(N_{U},\; t\right)\right]\geq\eta\right\} ,\label{eq:determine N2-1}
\end{equation}
\begin{equation}
N_{ST}^{*}=\min_{N_{ST}}\left\{ E\left[P\left(N_{U}+1,\; t\right)\right]-E\left[P\left(N_{U},\; t\right)\right]\leq\gamma\right\} ,\label{eq:determine N2-2}
\end{equation}
where $\eta$ is a predetermined probability and $\gamma$ is a threshold. }

\textit{We denote by $N_{IT}$ the selected number of transmit antennas
on IAP. $N_{IT}$ is determined by }

\textit{
\begin{equation}
N_{IT}=N_{R}Q'\left(N_{1}+\frac{N_{U}^{*}S_{A_{1}\cap A_{2}}}{S_{A_{2}}}\right).\label{eq:NIT-1}
\end{equation}
where $N_{U}^{*}=N_{ST}^{*}/N_{R}-N_{1}^{'}$ and $Q'$} indicates
the QoS requirements of\textit{ }inhabitants.

\section{Downlink Inter-sub-small Interference Control}

As discussed above, for our system, due to the use of BD, no inter-sub-small
cell interference is caused by downlink transmission under the situation
that IAP and SAP acquire perfect IAP-to-sojourner CSI and SAP-to-inhabitant
CSI respectively. However, the interference inflicted by SSSC on ISSC
has its origins in BD precoding exploiting CSI with inevitable uncertainty.
We proposed an interference control scheme named \textit{BDBF} for
secondary systems which are hampered from perfect secondary-to-primary
CSI. \textit{BDBF} utilizes BD Precoding based on uncertain CSI in
conjunction with auxiliary optimal BF. In view of the adoption of
OFDM, we turn the design of auxiliary optimal BF of \textit{BDBF}
into a multi-user multi-subcarrier optimal BF design under channel
uncertainties. For such optimization problems that contain the uncertainty
region constraint, S-Procedure is an efficacious tool to transform
semi-infinite programs into equivalent semi-definite programs \cite{Yu Zhang Distributed Optimal Beamformers for Cognitive Radios Robust to Channel Uncertainties,T. Al-Khasib Dynamic}.
We prove and verify that \textit{BDBF} performs better than the interference
controller using optimal BF alone for gaining more capacity within
a bearably large radius of uncertainty region.

\subsection{Block Diagonalization}

For each user $u_{i}$, BD algorithm \cite{BD} constructs a precoding
matrix $\mathbf{W}_{t,u_{i},s}$ which achieves the zero-interference
constraint, i.e. 
\begin{equation}
\mathbf{H}_{t,u_{j},s}\mathbf{W}_{t,u_{i},s}=0,\,\forall j\neq i,\label{eq:zero interference}
\end{equation}
where $\mathbf{H}_{t,u_{j},s}\in N_{r}\times N_{t}$ denotes the channel
matrix from transmitter $t$ to user $u_{j}$ at subcarrier $s$,
$N_{r}$ denotes the number of receive antennas at every UE and $N_{t}$
denotes the number of transmit antennas at transmitter $t$. 

Suppose that $N_{u}$ users are within the coverage area of transmitter
$t$ and we define 
\begin{equation}
\tilde{\mathbf{H}}_{t,u_{i},s}=\left[\mathbf{H}_{t,u_{1},s}^{T}\ldots\mathbf{H}_{t,u_{i-1},s}^{T}\;\mathbf{H}_{t,u_{i+1},s}^{T}\ldots\mathbf{H}_{t,u_{N_{u}},s}^{T}\right]^{T}.\label{eq:H}
\end{equation}
Hereupon, Equation (\ref{eq:zero interference}) has another equivalent
form

\begin{equation}
\tilde{\mathbf{H}}_{t,u_{i},s}\mathbf{W}_{t,u_{i},s}=0.\label{eq:another form}
\end{equation}
From Equation (\ref{eq:another form}), it is deduced that $\mathbf{W}_{t,u_{i},s}$
is a basis set in the null space of $\tilde{\mathbf{H}}_{t,u_{i},s}$.
The Singular Value Decomposition (SVD) of $\tilde{\mathbf{H}}_{t,u_{i},s}$
is expressed as 
\begin{equation}
\tilde{\mathbf{H}}_{t,u_{i},s}=\mathbf{\tilde{U}}_{t,u_{i},s}\left[\tilde{\boldsymbol{\Lambda}}_{t,u_{i},s}\mathbf{0}\right]\left[\mathbf{\tilde{V}}_{t,u_{i},s}^{\left(1\right)}\mathbf{\tilde{V}}_{t,u_{i},s}^{\left(0\right)}\right]^{\mathcal{H}},\label{eq:SVD}
\end{equation}
where $\mathbf{\tilde{V}}_{t,u_{i},s}^{\left(0\right)}\in\mathbb{C^{\mathrm{\mathit{\mathrm{N_{t}\times\left(N_{t}-rank\left(\mathbf{\tilde{H}}_{\mathit{t,u_{i},s}}\right)\right)}}}}}$
contains the right-singular vectors corresponding to the zero singular
values of $\mathbf{\tilde{H}}_{t,u_{i},s}$, i.e. $\mathbf{\tilde{V}}_{t,u_{i},s}^{\left(0\right)}$
forms a null space basis of $\mathbf{\tilde{H}}_{t,u_{i},s}$. Therefore
$\mathbf{W}_{t,u_{i},s}$ can be constructed by any linear combinations
of the columns in $\mathbf{\tilde{V}}_{t,u_{i},s}^{\left(0\right)}$.
Correspondingly, $\mathbf{H}_{t,u_{i},s}\mathbf{\tilde{V}}_{t,u_{i},s}^{\left(0\right)}$
forms the equivalent channel.

The SVD precoding can be performed using the equivalent channel:

\begin{equation}
\mathbf{H}_{t,u_{i},s}\mathbf{\tilde{V}}_{t,u_{i},s}^{\left(0\right)}=\mathbf{U}_{t,u_{i},s}\left[\begin{array}{cc}
\boldsymbol{\Lambda}_{t,u_{i},s} & \mathbf{0}\\
\mathbf{0} & \mathbf{0}
\end{array}\right]\left[\mathbf{V}_{t,u_{i},s}^{\left(1\right)}\mathbf{V}_{t,u_{i},s}^{\left(0\right)}\right]^{\mathcal{H}},\label{eq:SVD2}
\end{equation}
where $\mathbf{V}_{t,u_{i},s}^{\left(1\right)}\in\mathbb{C^{\mathrm{\mathit{\mathrm{\left(N_{t}-rank\left(\mathbf{\tilde{H}}_{\mathit{t,u_{i},s}}\right)\right)\times rank\left(\mathbf{H}_{t,u_{i},s}\mathbf{\tilde{V}}_{t,u_{i},s}^{\left(0\right)}\right)}}}}}$ contains
the right-singular vectors corresponding to the non-zero singular
values of $\mathbf{H}_{t,u_{i},s}\mathbf{\tilde{V}}_{t,u_{i},s}^{\left(0\right)}$.
$\mathbf{V}_{t,u_{i},s}^{\left(1\right)}$ is used as SVD precoder
and $\mathbf{U}_{t,u_{i},s}^{\mathcal{H}}$ is used as SVD decoder.
Consequently, the precoding matrix $\mathbf{W}_{t,u_{i},s}$ can be
chosen as $\mathbf{\tilde{V}}_{t,u_{i},s}^{\left(0\right)}\mathbf{V}_{t,u_{i},s}^{\left(1\right)}$.

For the full rank transmission, i.e. the number of streams sent to
a UE is no less than the number of receive antennas, it is needed
that $N_{t}-\mathrm{rank}\left(\tilde{\mathbf{H}}_{t,u_{i},s}\right)\geq N_{r}$,
i.e. $N_{t}\geq N_{r}\mathit{N}_{u}$. In this paper, we consider
the situation that the number of streams sent to a UE equals the number
of receive antennas, i.e $N_{t}=N_{r}N_{u}$. Accordingly, $\mathbf{W}_{t,u_{i},s}$
is a matrix of size $N_{t}\times N_{r}$.

\subsection{\textit{BDBF}}

As explained in Remark 1, SAP-to-inhabitant CSI is unable to be perfect.
The SAP-to-inhabitant channel matrix at subcarrier $s$ can be expressed
as

\begin{equation}
\mathbf{H}_{SAP,I_{i},s}=\hat{\mathbf{H}}_{SAP,I_{i},s}+\mathbf{\triangle\mathbf{H}_{\mathit{SAP,I_{i},s}}},\label{eq:CSI}
\end{equation}
where $\hat{\mathbf{H}}_{SAP,I_{i},s}$ is the estimated channel matrix
from transmitter $SAP$ to user $I_{i}\in U_{I}^{'}$ at subcarrier
$s$ and $\mathbf{\triangle\mathbf{H}_{\mathit{SAP,I_{i},s}}}$ is
the channel uncertainty matrix from transmitter $SAP$ to user $I_{i}\in U_{I}^{'}$
at subcarrier $s$. In this paper, we adopt a channel uncertainty
model \cite{Robust cognitive beamforming with bounded channel uncertainties}
which defines the uncertainty region as

\begin{eqnarray}
\triangle\left(\epsilon\right)=\left\{ \triangle\mathbf{H}_{\mathit{SAP,I_{i},s}}\mid\mathrm{Tr}\left\{ \mathbf{E}\right\} \leq\epsilon^{2}\right\} ,\label{eq:uncertainty region}
\end{eqnarray}
where $\mathbf{E=\triangle\mathbf{H}_{\mathit{SAP,I_{i},s}}}\mathbf{\mathbf{\triangle\mathbf{H}_{\mathit{SAP,I_{i},s}}^{\mathit{\mathcal{H}}}}}$.

\begin{proposition}
For secondary systems which are hampered from perfect secondary-to-primary
CSI, secondary systems can exploit \textit{BDBF}: BD precoding based on uncertain
CSI in conjunction with auxiliary optimal BF to control interference
to primary systems.

After \textit{BDBF}, the transmitted signal from SAP to sojourner $S_{i}$
can be expressed as

\begin{equation}
\mathbf{S}_{SAP,S_{i},s}=\mathbf{W}_{SAP,S_{i},s}\mathbf{P}_{SAP,S_{i},s}\mathbf{x}_{SAP,S_{i},s}\label{eq:BDBF}
\end{equation}
where
\end{proposition}
{\small $\mathbf{x}_{SAP,I_{i},s}\in\mathbb{C}^{N_{ST}^{*}\times1}$}
\textit{\small is the information symbol vector transmitted from SAP
to }{\small $I_{i}$} \textit{\small at subcarrier} {\small $s$}
\textit{\small with covariance matrix} {\small $\mathbb{E}\left\{ \mathbf{x}_{SAP,I_{i},s}\mathbf{x}_{SAP,I_{i},s}^{\mathcal{H}}\right\} =\mathbf{I}_{N_{ST}^{*}}$}\textit{\small ,}{\small \par}

{\small $\mathbf{P}_{SAP,I_{i},s}\in\mathbb{C}^{N_{R}\times N_{ST}^{*}}$}
\textit{\small is the auxiliary optimal BF, }{\small \par}

\textit{\small and} {\small $\mathbf{W}_{SAP,S_{i},s}$} \textit{\small is
the BD precoding matrix of size} {\small $N_{ST}^{*}\times N_{R}$.}{\small \par}

{\small $\mathbf{W}_{SAP,S_{i},s}$ }\textit{\small is generated by}
{\footnotesize $\tilde{\mathbf{H}}_{SAP,S_{i},s}=\left[\breve{\mathbf{H}}_{SAP,S_{i},s}\;\mathbf{H}_{SAP,I_{1},s}^{T}\ldots\mathbf{H}_{SAP,I_{N_{1}},s}^{T}\right]^{T}$}{\small{}
}\textit{\small and} {\footnotesize $\breve{\mathbf{H}}_{SAP,S_{i},s}=\left[\mathbf{H}_{SAP,S_{1},s}^{T}\ldots\mathbf{H}_{SAP,S_{i-1},s}^{T}\;\mathbf{H}_{SAP,S_{i+1},s}^{T}\ldots\mathbf{H}_{SAP,S_{N_{2}\left(t\right)},s}^{T}\right]$.} 

For inhabitant $I_{i}$, the received signal at subcarrier
$s$ is given by

\begin{eqnarray}
\mathbf{y}_{I_{i},s} & = & \mathbf{U}_{IAP,I_{i},s}^{\mathcal{H}}\mathbf{H}_{IAP,I_{i},s}\mathbf{W}_{IAP,I_{i},s}\mathbf{x}_{IAP,I_{i},s}\nonumber \\
 & + & \sum_{S_{i}\in U_{S}}\mathbf{H}_{SAP,I_{i},s}\mathbf{W}_{SAP,S_{i},s}\mathbf{P}_{SAP,S_{i},s}\mathbf{x}_{SAP,S_{i},s}\nonumber \\
 & + & \mathbf{n}_{I_{i},s},\label{eq:received signal}
\end{eqnarray}
where 

{\small $\mathbf{x}_{IAP,I_{i},s}\in\mathbb{C}^{N_{IT}\times1}$ is
the information symbol vector transmitted from $IAP$ to $I_{i}$
at subcarrier $s$ with covariance matrix $\mathbb{E}\left\{ \mathbf{x}_{IAP,I_{i},s}\mathbf{x}_{IAP,I_{i},s}^{\mathcal{H}}\right\} =\mathbf{I}_{N_{IT}}$,}{\small \par}

{\small $\mathbf{n}_{I_{i},s}\in\mathbb{C}^{N_{R}\times1}$ is the
i.i.d. circularly symmetric white Gaussian noise at subcarrier $s$
at reciever $I_{i}$ with covariance matrix $\mathbb{E}\left\{ \mathbf{n}_{I_{i},s}\mathbf{n}_{I_{i},s}^{\mathcal{H}}\right\} =\mathbf{I}_{N_{R}}$,}{\small \par}

{\small $\mathbf{W}_{IAP,I_{i},s}$ is the BD precoding matrix of
size $N_{IT}\times N_{R}$,}{\small \par}

{\small and $\mathbf{U}_{IAP,I_{i},s}^{\mathcal{H}}$ is the SVD decoding
matrix of size $N_{R}\times N_{R}$.}{\small \par}

{\small $\mathbf{W}_{IAP,I_{i},s}$ and $\mathbf{U}_{IAP,I_{i},s}^{\mathcal{H}}$
are generated by }\textit{\small $\tilde{\mathbf{H}}_{IAP,I_{i},s}=\left[\mathbf{\overline{H}}_{IAP,I_{i},s}\;\mathbf{H}_{IAP,S_{1},s}^{T}\ldots\mathbf{H}_{IAP,S_{N_{2}(t)},s}^{T}\right]^{T}$
}{\small and }\textit{\small $\mathbf{\mathbf{\overline{H}}}_{IAP,I_{i},s}=\left[\mathbf{H}_{IAP,I_{1},s}^{T}\ldots\mathbf{H}_{IAP,I_{i-1},s}^{T}\;\mathbf{H}_{IAP,I_{i+1},s}^{T}\ldots\mathbf{H}_{IAP,I_{N_{1}},s}^{T}\right]$}{\small .}{\small \par}

\begin{theorem}
The auxiliary optimal BF of \textit{BDBF} can be formulated as the following
multi-user multi-subcarrier optimal BF design under channel uncertainties:
trying for the maximum capacity of secondary system under the precondition
that the interference inflicted to the primary system is kept below
the threshold.
\end{theorem}

\begin{subequations}
\begin{equation}
\left(\mathbf{\mathrm{\mathbf{P1}}}\right)\;\max_{\boldsymbol{Q}_{SAP,S_{i},s}}\sum_{s=1}^{N_{C}}\sum_{S_{i}\in U_{S}}\log\mathrm{det}\left(\mathbf{\mathbf{I}_{\mathit{N_{R}}}}+\frac{SNR}{N_{ST}^{*}}\mathbf{A}\right)\label{eq:P1a}
\end{equation}

\begin{equation}
\mathrm{s.t.}\;\boldsymbol{Q}_{SAP,S_{i},s}\geqslant0,\;\forall S_{i}\in U_{S},\;\forall s,\label{eq:P1b}
\end{equation}

\begin{equation}
\sum_{S_{i}\in U_{S}}\mathrm{Tr}\left(\boldsymbol{\breve{Q}}_{SAP,S_{i},s}\right)\leq P_{SAP},\;\forall S_{i}\in U_{S},\;\forall s,\label{eq:P1c}
\end{equation}

\begin{equation}
\mathrm{Tr}\left(\mathbf{B}\right)\leq\zeta,\;\forall\triangle\mathrm{\mathbf{H}}_{SAP,I_{i},s}\in\triangle\left(\epsilon\right),\;\forall S_{i}\in U_{S},\;\forall I_{i}\in U_{I}^{'},\;\forall s,\label{eq:P1d}
\end{equation}
\end{subequations}

\textit{where}

\textit{\small $\boldsymbol{Q}_{SAP,S_{i},s}=\mathbf{P}_{SAP,S_{i},s}\mathbf{P}_{SAP,S_{i},s}^{\mathcal{H}},$}{\small \par}

\textit{\small $\boldsymbol{\breve{Q}}_{SAP,S_{i},s}=\mathbf{W}_{SAP,S_{i},s}\boldsymbol{Q}_{SAP,S_{i},s}\mathbf{\mathbf{W}_{\mathit{SAP,S_{i},s}}^{\mathcal{H}}},$}{\small \par}

\textit{\small $\mathbf{A}=\boldsymbol{\Lambda}_{SAP,S_{i},s}\boldsymbol{Q}_{SAP,S_{i},s}\mathrm{\mathbf{\boldsymbol{\Lambda}}}_{SAP,S_{i},s}^{\mathcal{H}},$}{\small \par}

\textit{\small $\mathbf{B}=\mathrm{\mathbf{H}}_{SAP,I_{i},s}\boldsymbol{\breve{Q}}_{SAP,S_{i},s}\mathrm{\mathbf{H}}_{SAP,I_{i},s}^{\mathcal{H}},$}{\small \par}

\textit{$P_{SAP}$ is the maximum downlink transmit power of SAP, }

\textit{and $\zeta$ is the inter-sub-small interference tolerable
threshold on each subcarrier for inhabitant $I_{i}\in U_{I}^{'}$.}

\textit{$\boldsymbol{Q}_{SAP,S_{i},s}$ is a positive semidefinite
matrix to render certain that the elements on the main diagonal of
$\mathbf{A}$ and $\mathbf{B}$ are real and non-negative. Additionally,
$\boldsymbol{Q}_{SAP,S_{i},s}$ is a symmetric matrix to ensure that
linear matrix inequality (LMI) holds.}

$\triangle\left(\epsilon\right)$ is a set of infinite cardinality
and accordingly the inequality (\ref{eq:P1d}) imposes infinite constraints.
This means that $\left(\mathrm{P1}\right)$ is a semi-infinite program.
$\left(\mathrm{P1}\right)$ can be transformed into an equivalent
semi-definite program $\left(\mathrm{P2}\right)$ with the constraint
(\ref{eq:P2d}) which contains finite constraints instead of the constraint
(\ref{eq:P1d}):

\begin{subequations}
\begin{equation}
\left(\boldsymbol{\mathbf{\mathbf{\mathrm{P2}}}}\right)\;\max_{\boldsymbol{Q}_{SAP,S_{i},s}}\sum_{s=1}^{N_{C}}\sum_{S_{i}\in U_{S}}\log\mathrm{det}\left(\mathbf{\mathbf{I}_{\mathit{N_{R}}}}+\frac{SNR}{N_{ST}^{*}}\mathbf{A}\right)\label{eq:P2a}
\end{equation}

\begin{equation}
\mathrm{s.t.}\;\boldsymbol{Q}_{SAP,S_{i},s}\geqslant0,\;\forall S_{i}\in U_{S},\;\forall s,\label{eq:P2b}
\end{equation}

\begin{equation}
\sum_{S_{i}\in U_{S}}\mathrm{Tr}\left(\boldsymbol{\breve{Q}}_{SAP,S_{i},s}\right)\leq P_{SAP},\;\forall S_{i}\in U_{S},\;\forall s,\label{eq:P2c}
\end{equation}

\begin{eqnarray}
\left[\begin{array}{cc}
\alpha\mathbf{\mathbf{I}}_{N_{T}\times N_{R}}+\mathbf{G} & \mathbf{J}^{\mathcal{H}}\\
\mathbf{J} & k-\alpha\epsilon^{2}
\end{array}\right] & \geq & 0,\nonumber \\
\forall\alpha\geq0,\;\forall I_{i}\in U_{I}^{'},\;\forall S_{i}\in U_{S},\;\forall s\label{eq:P2d}
\end{eqnarray}
\end{subequations}

\begin{IEEEproof}
According to the relationship between the trace operator and the vec
operator $\mathrm{Tr}\left(\mathbf{C^{\mathcal{H}}DC}\right)=\mathrm{vec}\left(\mathbf{C}\right)^{\mathcal{H}}\left(\mathbf{I}\otimes\mathbf{D}\right)\mathrm{vec}\left(\mathbf{C}\right)$
and $\mathrm{Tr}\left(\mathbf{C^{\mathcal{H}}D}\right)=\mathrm{vec}\left(\mathbf{C}\right)^{\mathcal{H}}\mathrm{vec}\left(\mathbf{D}\right)$,
the inequality (\ref{eq:P1d}) can be converted to an equivalent form:

\textit{\small 
\begin{eqnarray}
-vec\left(\mathbf{\mathrm{\triangle}\mathbf{H}_{\mathit{SAP,I_{i},s}}^{\mathit{\mathcal{H}}}}\right)^{\mathit{\mathcal{H}}}\left(\mathbf{I}_{N_{R}}\otimes\boldsymbol{\breve{Q}}_{SAP,S_{i},s}\right)vec\left(\mathbf{\mathrm{\triangle}\mathbf{H}_{\mathit{SAP,I_{i},s}}^{\mathit{\mathcal{H}}}}\right)\nonumber \\
-2\mathfrak{R}\left(\mathrm{vec}\left(\mathbb{\mathbf{\boldsymbol{\breve{Q}}_{\mathit{SAP,S_{i},s}}}^{\mathit{\mathcal{H}}}}\mathbf{\mathbf{\hat{H}}_{\mathit{SAP,I_{i},s}}^{\mathit{\mathcal{H}}}}\right)^{\mathcal{H}}\mathrm{vec}\left(\mathrm{\triangle}\mathbf{H}_{\mathit{SAP,I_{i},s}}^{\mathit{\mathcal{H}}}\right)\right)\nonumber \\
-\mathrm{Tr}\left(\mathbf{\mathbf{\hat{H}}_{\mathit{SAP,I_{i},s}}\mathbf{\boldsymbol{\breve{Q}}_{\mathit{SAP,S_{i},s}}}}\mathbf{\mathbf{\hat{H}}_{\mathit{SAP,I_{i},s}}^{\mathit{\mathcal{H}}}}\right)\nonumber \\
+\zeta\geq0,\nonumber \\
\forall vec\left(\mathbf{\mathrm{\mathrm{\triangle}}\mathbf{H}_{\mathit{SAP,I_{i},s}}^{\mathit{\mathcal{H}}}}\right)^{\mathit{\mathcal{H}}}vec\left(\mathbf{\mathrm{\triangle}\mathbf{H}_{\mathit{SAP,I_{i},s}}^{\mathit{\mathcal{H}}}}\right)\leq\epsilon^{2},\nonumber \\
\forall I_{i}\in U_{I}^{'},\;\forall S_{i}\in U_{S},\;\forall s\label{eq:trace vec}
\end{eqnarray}
}{\small \par}

The result of S-Procedure \cite{S-procedure,Convex optimization}
presents a condition that makes the infinite constraints imposed by
the inequality (\ref{eq:trace vec}) hold: if and only if there exists
$\alpha\geqslant0$ such that the inequality (\ref{eq:condition})
is ture. 

\begin{equation}
\left[\begin{array}{cc}
\alpha\mathbf{\mathbf{I}}_{N_{T}\times N_{R}}+\mathbf{G} & \mathbf{J}^{\mathcal{H}}\\
\mathbf{J} & k-\alpha\epsilon^{2}
\end{array}\right]\geq0,\;\forall I_{i}\in U_{I}^{'},\;\forall S_{i}\in U_{S},\;\forall s\label{eq:condition}
\end{equation}
where $\mathbf{G}=-\mathbf{\mathbf{I}}_{N_{R}}\otimes\boldsymbol{\breve{Q}}_{SAP,S_{i},s}$,
$\mathbf{J}=-\mathrm{vec}\left(\mathbb{\mathbf{\boldsymbol{\breve{Q}}}_{\mathit{SAP,S_{i},s}}^{\mathit{\mathcal{H}}}}\mathbf{\mathbf{\hat{H}}_{\mathit{SAP,I_{i},s}}^{\mathit{\mathcal{H}}}}\right)$,
and $k=\zeta-\mathrm{Tr}\left(\mathbf{\mathbf{\hat{H}}_{\mathit{SAP,I_{i},s}}}\boldsymbol{\breve{Q}}_{SAP,S_{i},s}\mathbf{\mathbf{\hat{H}}_{\mathit{SAP,I_{i},s}}^{\mathit{\mathcal{H}}}}\right)$. 
\end{IEEEproof}
The semi-definite program $\left(\mathrm{P2}\right)$ can be solved
by interior point methods \cite{Convex optimization,CVX}.
\begin{theorem}
\textit{BDBF} can gain more capacity than the interference controller
that using optimal BF alone within a bearably large radius of uncertainty
region.\end{theorem}
\begin{IEEEproof}
We denote by $\mathbf{W}_{SAP,S_{i},s}^{a}$ and $\mathbf{W}_{SAP,S_{i},s}^{b}$
the BD precoding matrix which is generated by \textit{\small $\tilde{\mathbf{H}}_{SAP,S_{i},s}^{a}=\left[\breve{\mathbf{H}}_{SAP,S_{i},s}\;\mathbf{\hat{H}}_{IAP,I_{1},s}^{T}\ldots\mathbf{\hat{H}}_{IAP,I_{N_{1}},s}^{T}\right]^{T}$}
and the BD precoding matrix which is generated by \textit{\small $\tilde{\mathbf{H}}_{SAP,S_{i},s}^{b}=\left[\breve{\mathbf{H}}_{SAP,S_{i},s}\right]^{T}$},
respectively. Since $\mathbf{W}_{SAP,S_{i},s}^{a}$ is a basis set
in the null space of $\tilde{\mathbf{H}}_{SAP,S_{i},s}^{a}$, then
for $\epsilon^{2}\leq\iota$, we have \textit{\small $\mathrm{Tr}\left(\mathrm{\mathbf{H}}_{SAP,I_{i},s}\mathbf{W}_{SAP,S_{i},s}^{a}\mathbf{W}_{SAP,S_{i},s}^{\mathcal{\mathit{a}H}}\mathrm{\mathbf{H}}_{SAP,I_{i},s}^{\mathcal{H}}\right)<\mathrm{Tr}\left(\mathrm{\mathbf{H}}_{SAP,I_{i},s}\mathbf{W}_{SAP,S_{i},s}^{b}\mathbf{W}_{SAP,S_{i},s}^{\mathcal{\mathit{b}H}}\mathrm{\mathbf{H}}_{SAP,I_{i},s}^{\mathcal{H}}\right)$},
where $\iota$ is a constant. Accordingly, for $\epsilon^{2}\leq\iota$,
when $\mathrm{Tr}\left(\mathbf{B}^{a}\right)=\mathrm{Tr}\left(\mathbf{B}^{b}\right)=\zeta$,
we have \textit{\small $\mathrm{Tr}\left(\boldsymbol{Q}_{SAP,S_{i},s}^{a}\right)>\mathrm{Tr}\left(\boldsymbol{Q}_{SAP,S_{i},s}^{b}\right)$},
where \textit{\small $\mathbf{B}^{a}=\mathrm{\mathbf{H}}_{SAP,I_{i},s}\mathbf{W}_{SAP,S_{i},s}^{a}\boldsymbol{Q}_{SAP,S_{i},s}^{a}\mathbf{W}_{SAP,S_{i},s}^{\mathcal{\mathit{a}H}}\mathrm{\mathbf{H}}_{SAP,I_{i},s}^{\mathcal{H}}$}
and \textit{\small $\mathbf{B}^{b}=\mathrm{\mathbf{H}}_{SAP,I_{i},s}\mathbf{W}_{SAP,S_{i},s}^{b}\boldsymbol{Q}_{SAP,S_{i},s}^{b}\mathbf{W}_{SAP,S_{i},s}^{\mathcal{\mathit{b}H}}\mathrm{\mathbf{H}}_{SAP,I_{i},s}^{\mathcal{H}}$}.
Therefore, for $\epsilon^{2}\leq\iota$, based upon Theorem 1., the
capacity of SSSC using $\mathbf{W}_{SAP,S_{i},s}^{a}$ is larger than
that of SSSC using $\mathbf{W}_{SAP,S_{i},s}^{b}$. 
\end{IEEEproof}
We simulate the SSSC using $\mathbf{W}_{SAP,S_{i},s}^{a}$ and $\mathbf{W}_{SAP,S_{i},s}^{b}$
for different $\epsilon^{2}$ in Section IV. We find that $\iota$
can be much larger than the normally acceptable radius of uncertainty
region.

\section{Simulation}

\subsection{Number of Transmit Antennas on Sub-small Cell Access Points}

Since the number of transmit antennas on SAP and that on IAP are in
direct relation with the number of concurrent sojourners, we design
8 data sets to observe and analyze the number of concurrent sojourners
in 8 representative circumstances.

We consult a medium-sized software enterprise in Paris to help us
to design data sets to approximate the actual situation. We receive
the following information on visitors in this enterprise:
\begin{itemize}
\item There are more than fifty visitors daily at an average.
\item In view of long term observation, there is no such a working day when
the number of visitors is particularly large or small within five
working days.
\item After the holidays, the number of visitors is inclined to peak.
\end{itemize}

First, we exclude those days when the number of visitors is inclined
to peak after holidays. Based on the second item above, we consider
that the observations of the number of arrivals at the reference time
on different working days are independent and identically distributed
(i.i.d.). We observe the sample mean of the cumulative probability
of the number of concurrent sojourners during one entire working day
based on the first six data sets. Here, we set $Q=1$. We adopt $E\left[P\left(N_{U},\; t\right)\right]$
which is defined in Equation (\ref{eq:mean value of probability})
to indicate the sample mean of the cumulative probability of the number
of concurrent sojourners during the observation time interval. We
interpret $P\left(N_{U},\; t\right)$ as the probability of the number
of concurrent sojourners being less than $N_{U}$ at time $t$. 

Then we consider separately those days after holidays. We also consider
that the observations of the number of arrivals at the reference time
on different working days during those days are i.i.d.. We observe
the sample mean of the cumulative probability of the number of concurrent
sojourners during one entire working day based on the last two data
sets.

We list the attributes of these 8 test data sets in Table (\ref{tab:attributes}).

\begin{table}
\caption{Attributes of test data sets\label{tab:attributes}}
\begin{tabular}{|c|c|c|c|}
\hline 
\multicolumn{4}{|c|}{{\scriptsize Data set index}}\tabularnewline
\hline 
\multicolumn{4}{|c|}{{\scriptsize Observation time interval (9 am to 5 pm)}}\tabularnewline
\hline 
\multicolumn{4}{|c|}{{\scriptsize Total number of sojourners during the observation time
interval}}\tabularnewline
\hline 
\multicolumn{4}{|c|}{{\scriptsize Sample mean of the duration of stay of one sojourner
(minute)}}\tabularnewline
\hline 
\multicolumn{4}{|c|}{{\scriptsize Standard deviation of the duration of stay of one sojourner}}\tabularnewline
\hline 
\multicolumn{4}{|c|}{{\scriptsize Arrival rate (arrivals per time slot)}}\tabularnewline
\hline 
\hline 
{\scriptsize Data set 1} & {\scriptsize Data set 2} & {\scriptsize Data set 3} & {\scriptsize Data set 4}\tabularnewline
\hline 
{\scriptsize -} & {\scriptsize -} & {\scriptsize -} & {\scriptsize -}\tabularnewline
\hline 
{\scriptsize 60} & {\scriptsize 60} & {\scriptsize 60} & {\scriptsize 60}\tabularnewline
\hline 
{\scriptsize 60 } & {\scriptsize 60} & {\scriptsize 90} & {\scriptsize 90}\tabularnewline
\hline 
{\scriptsize 3.7947} & {\scriptsize 148.0447} & {\scriptsize 3.7947} & {\scriptsize 148.0447}\tabularnewline
\hline 
{\scriptsize 60/48 } & {\scriptsize 60/48 } & {\scriptsize 60/48} & {\scriptsize 60/48}\tabularnewline
\hline 
\hline 
{\scriptsize Data set 5} & {\scriptsize Data set 6} & {\scriptsize Data set 7} & {\scriptsize Data set 8}\tabularnewline
\hline 
{\scriptsize -} & {\scriptsize -} & {\scriptsize -} & {\scriptsize -}\tabularnewline
\hline 
{\scriptsize 60} & {\scriptsize 60} & {\scriptsize 90} & {\scriptsize 90}\tabularnewline
\hline 
{\scriptsize 60} & {\scriptsize 60} & {\scriptsize 60} & {\scriptsize 60}\tabularnewline
\hline 
{\scriptsize 3.7947} & {\scriptsize 148.0447} & {\scriptsize 2.7809} & {\scriptsize 134.5144}\tabularnewline
\hline 
{\scriptsize 5(10-11 am), 30/42(else) } & {\scriptsize 5(10-11 am), 30/42(else) } & {\scriptsize 60/48} & {\scriptsize 60/48}\tabularnewline
\hline 
\end{tabular}
\end{table}

The staffs' working hours are from 9 am to 5 pm. The visitors arrive
randomly during working hours. The observation time interval is set
to be consistent with the working hours for all data sets. We divide
the observation time interval into $48$ time slots, each $10$ minutes
in length. Data set 1 and data set 2 have the same attributes except
the standard deviation of the duration of stay. These two data sets
are considered as the benchmark for comparison and designing the other
data sets. Data set 3 and data set 4 are designed by adding 30 minutes
to every sample of the duration of stay in data set 1 and set 2. Data
set 5 and data set 6 are designed by changing the immutable arrival
rate of data set 1 and data set 2 to 5 arrivals per time slot during
10-11 am and 30/42 arrivals per time slot during the other working
hours. Date set 7 and set 8 are designed by adding 30 to the total
number of sojourners during the observation time interval of data
set 1 and data set 2 and make the the standard deviation of the duration
of stay of data set 7 and data set 8 to be close to that of data set
1 and data set 2 respectively.

We exploit EMPHT algorithm \cite{EM algo fitting,EMpht} to fit the
phase type distribution to the data of duration of stay in each data
set.{\scriptsize{} }The EMPHT is an iterative maximum likelihood estimation
algorithm which outputs the $m\times1$ vector $\boldsymbol{\alpha}$
and $m\times m$ matrix $\mathbf{R}$. We chose the number of phases
as 4, i.e. $m=4$. We acquire these two parameters based on each data
set after sufficient iterations when the likelihood function reaches
zero growth and accordingly obtain the CDF of the duration of stay
for each data set.

\begin{figure}
\subfloat[\label{fig:2a}]{\includegraphics[width=8.5cm,height=5cm]{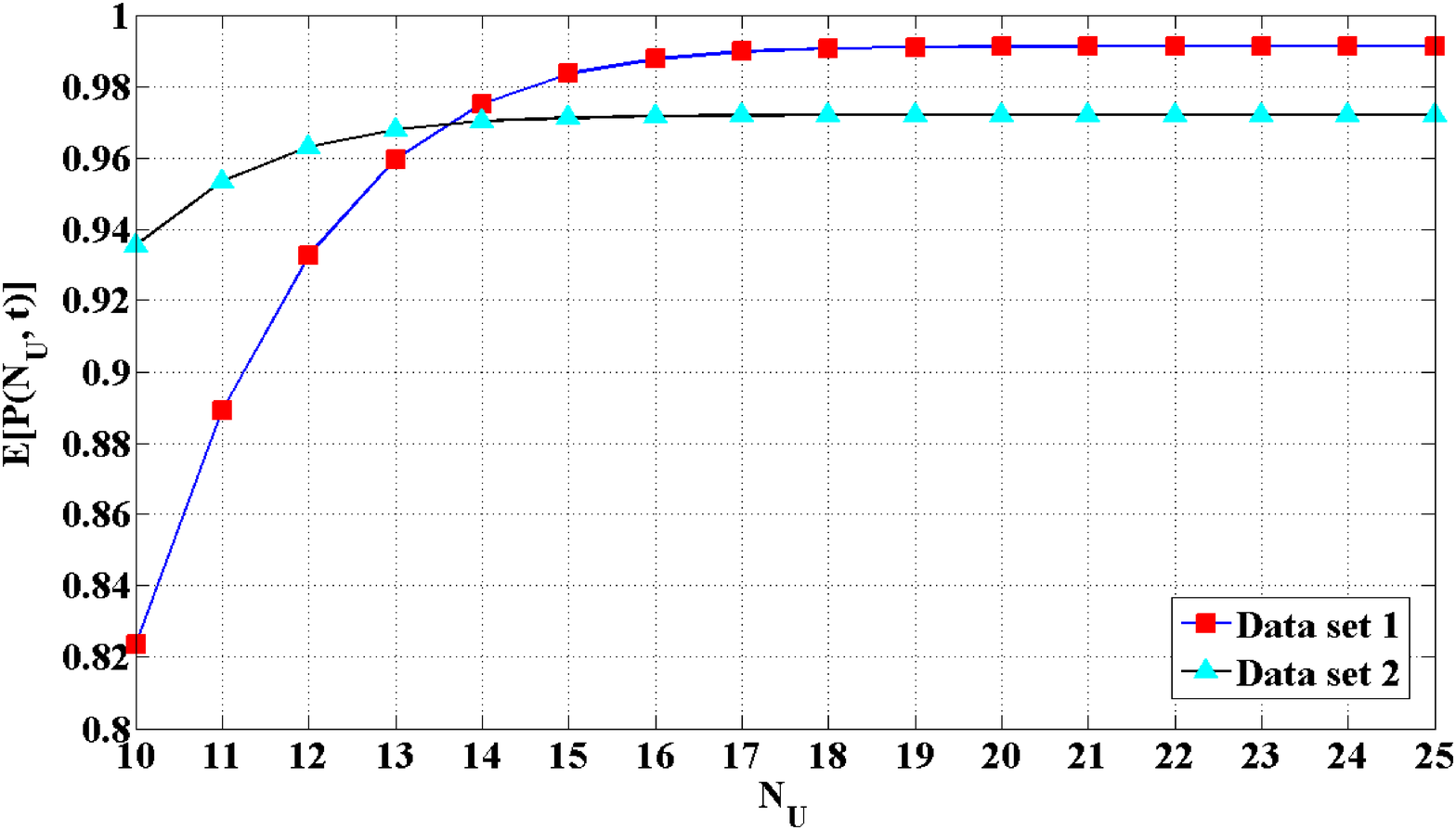}}

\subfloat[\label{fig:2b}]{\includegraphics[width=8.5cm,height=5cm]{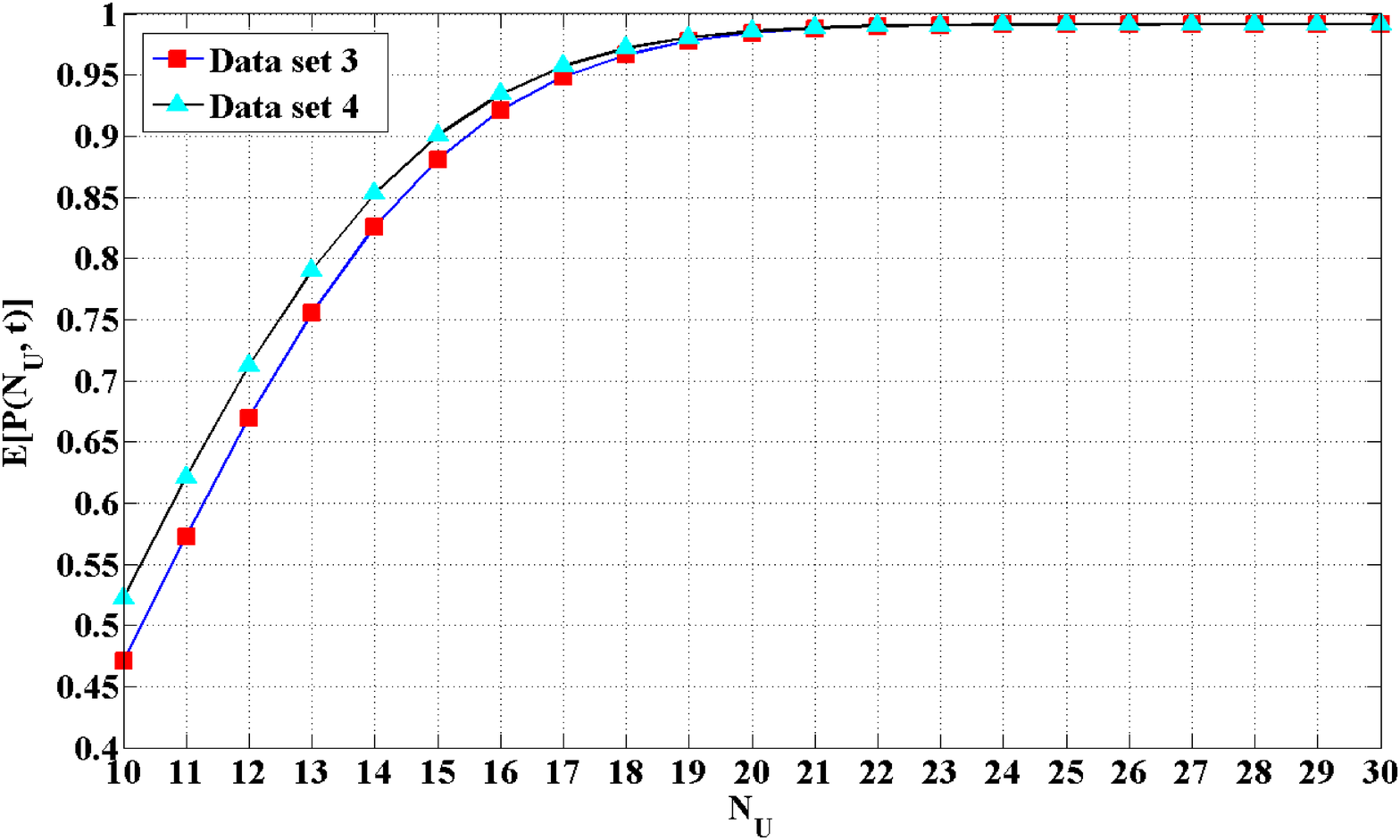}}

\subfloat[\label{fig:2c}]{\includegraphics[width=8.5cm,height=5cm]{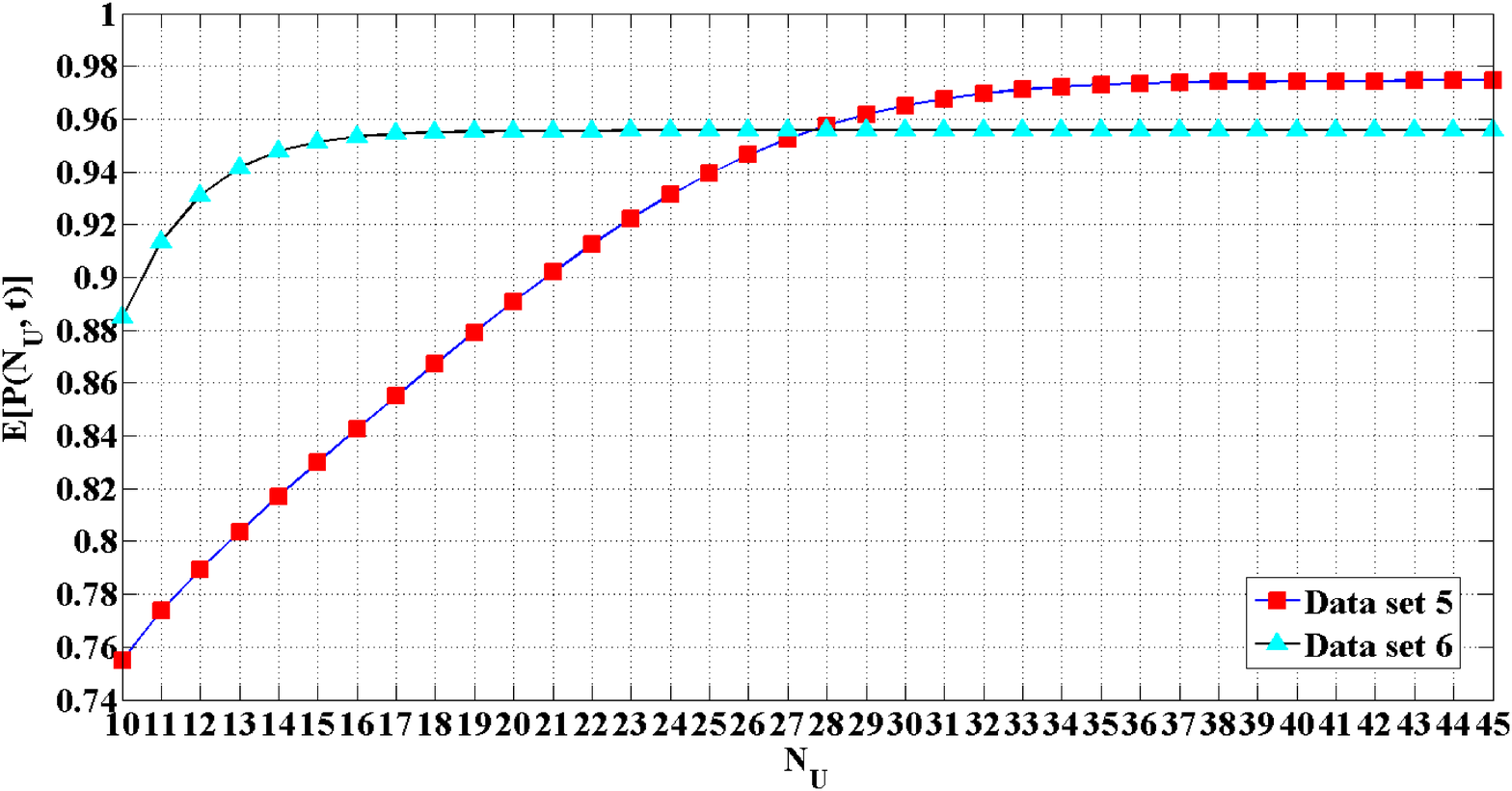}}

\subfloat[\label{fig:2d}]{\includegraphics[width=8.5cm,height=5cm]{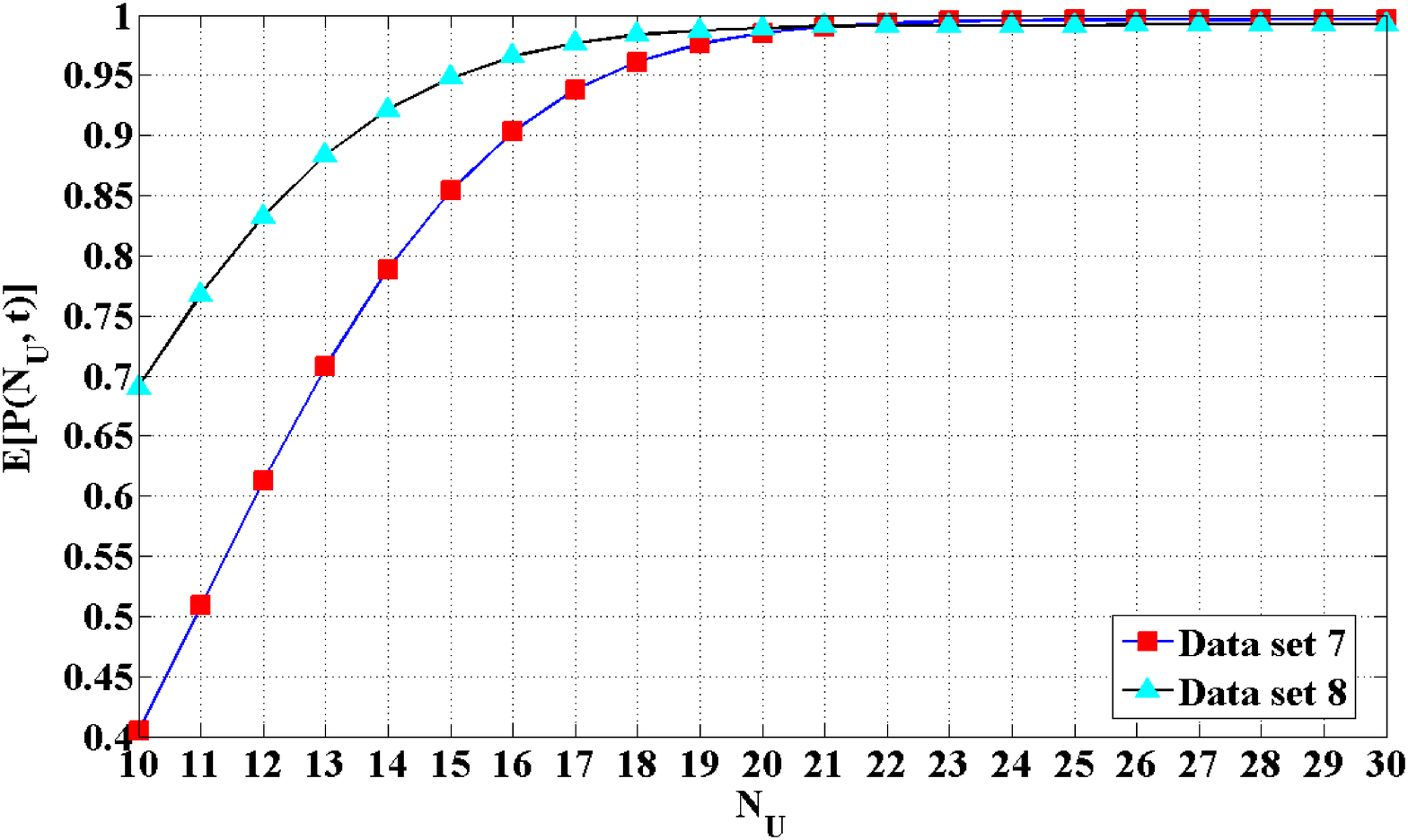}}\caption{Sample mean of the CDF of the number of concurrent sojourners during
the observation time interval base on each data set\label{fig:2}}
\end{figure}

We further calculate the mean value of the cumulative probability
of the number of concurrent sojourners during the observation time
interval base on each of these 8 data sets and showed them in Figure
\ref{fig:2}. From them, we can observe how the following factors
influence the number of concurrent sojourners and further impact the
number of transmit antennas on sub-small cell APs:
\begin{itemize}
\item the standard deviation of the duration of stay
\item the mean of the duration of stay
\item the distribution of arrival rate
\item the total number of sojourners during the observation time interval
\end{itemize}

From Figure \ref{fig:2}, we observe that as $N_{U}$ increases, its
sample mean of the cumulative probability of the number of concurrent
sojourners during the observation time interval based on 8 data sets
all exhibit uptrend and the rate of growth declines gradually. The
values of $E\left[P\left(N_{U}=10,\; t\right)\right]$ based on data
set 1 until data set 8 are: 0.8237, 0.9354, 0.4711, 0.5223, 0.7550,
0.8849, 0.4048 and 0.6905. The values of $E\left[P\left(N_{U}=20,\; t\right)\right]$
based on data set 1 until data set 8 are: 0.9911, 0.9719, 0.9841,
0.9857, 0.8908, 0.9554, 0.9852 and 0.9895. The pair of $n=\min_{N_{U}}\left\{ E\left[P\left(N_{U}+1,\; t\right)\right]-E\left[P\left(N_{U},\; t\right)\right]\leq0.001\right\} $
and $E\left[P\left(n,\; t\right)\right]$ based on data set 1 until
data set 8 are: $(19,\;0.991)$, $\left(15,\;0.9712\right)$, $\left(24,\;0.9912\right)$,
$\left(24,\;0.9911\right)$, $\left(38,\;0.9741\right)$, $\left(19,\;0.9553\right)$,
$\left(26,\;0.9961\right)$ and $\left(22,\;0.9912\right)$.

From the first two sets of experimental results, we can observe that:
the number of concurrent sojourners during the observation time interval
being more than 20 has a very small probability (<0.03) for all data
set 1, 2, 3, 4, 7 and 8 and a small probability (<0.11) for data set
5 and 6. This number being no more than 10 has a large probability
(>0.8) for data set 1 and 2, followed by data set 5 and 6 (>0.75).
For data set 3, 4, 7 and 8, This number being no more than 10 and
between 10 and 20 have almost the same probability ($\approx0.4$). 

Referring to Figure \ref{fig:2}, we can find that the increasement
of the mean of the duration of stay, non-uniform arrival rate or the
increase of the total number of sojourners during the observation
time interval improves the probability of the number of concurrent
sojourners being a relatively large number. The increasement of standard
deviation of the duration of stay brings down this probability. However,
when the mean of the duration of stay is relatively large, the standard
deviation of the duration of stay has minimal impact on this probability
as shown in Figure 2 (b). Accordingly, the number of transmit antennas
on sub-small cell access points are chosen to be relatively large
when the mean of the duration of stay or the total number of sojourners
is relatively large, the arrival rate is non-uniform or the standard
deviation of the duration of stay is relatively small. 

From the third set of experimental result, we can see that the augmentation
of standard deviation of the duration of stay causes that the convergence
of the sample means of the cumulative probability of the number of
concurrent sojourners during the observation time interval to be more
slowly and the low growth rate emergences earlier. But when the mean
of the duration of stay is relatively large, this phenomenon is not
obvious. So we need to choose a relatively small value for $\eta$
in Equation (\ref{eq:determine N2-1}) or $\gamma$ in Equation (\ref{eq:determine N2-2})
when the standard deviation of the duration of stay is relatively
large and the mean of the duration of stay is relatively small.

\subsection{BD precoding based on uncertain CSI in conjunction with auxiliary
optimal BF for downlink inter-sub-small interference control}

For simplicity we assume that the inter-sub-small cell interference
only exists in the overlapping area between ISSC and SSSC in our simulation. We consider that 2 sojourners and 2 inhabitants stay in the overlap
of ISSC and SSSC, where all the sojourners and inhabitants have the
same distance from the SAP. The path loss component is normalized
to one. $\mathrm{\mathbf{H}}_{SAP,S_{i},s}$ and $\mathbf{H}_{SAP,I_{i},s}$
are both modeled as a matrix with coefficients which are i.i.d. circularly
symmetric, complex Gaussian random variables, with zero mean and unit
variance. $\triangle\mathbf{H}_{\mathit{SAP,I_{i},s}}$ is modeled
as a matrix with coefficients which are i.i.d. circularly symmetric,
complex Gaussian random variables, with zero mean and variance $\sigma^{2}$.
We set that $\epsilon^{2}$ in Equation (\ref{eq:uncertainty region})
equals to $\sigma^{2}$. The number of receive antennas at every UE
is set to be 2 and the maximum Signal-to-Noise Ratio (SNR) at every
UE is set to be $20\,\mathrm{dB}$. The maximum downlink transmit
power of SAP is 10W. The SAP exploits 4 subcarriers for downlink transmission.
Each subcarrier occupies a normalized bandwidth of $1\,\mathrm{Hz}$.

\begin{figure}
\includegraphics[width=8.5cm,height=5cm]{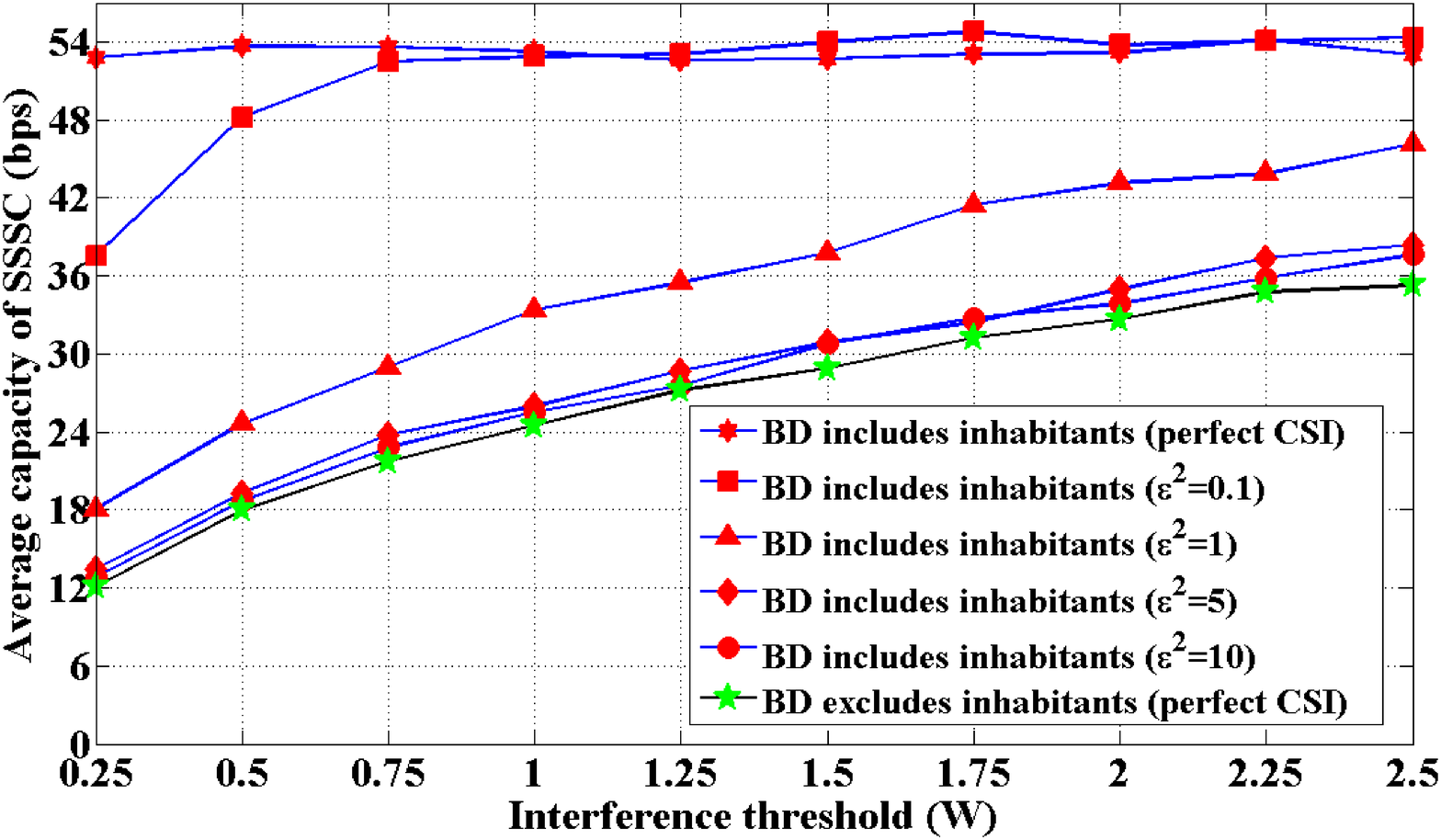}\caption{Average capacity of SSSC vs interference threshold\label{fig3:Average-capacity vs interference threshold}}
\end{figure}

\begin{figure}
\includegraphics[width=8.5cm,height=5cm]{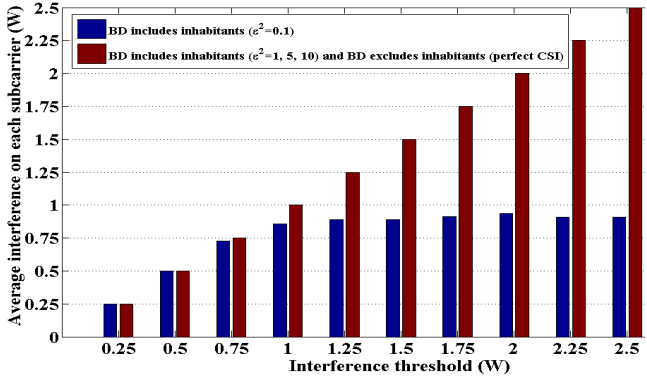}\caption{Average interference on each subcarrier vs. interference threshold\label{fig4:Average-interference vs Inter threshold}}
\end{figure}

For the purpose of comparing the performance of \textit{BDBF} with
the interference controller that uses optimal BF alone, we design
two test systems: 1. downlink BD precoding for the secondary sub-small
cell takes both sojourners and inhabitants who stay in the coverage
area into consideration, i.e. the SSSC using $\mathbf{W}_{SAP,S_{i},s}^{a}$;
2. downlink BD precoding for the secondary sub-small cell takes only
sojourners into consideration, i.e. the SSSC using $\mathbf{W}_{SAP,S_{i},s}^{b}$.
We simulate the above systems which exploit BD precoding based on
SAP-to-inhabitant CSI with different radius of uncertainty region
($\epsilon^{2}$=0, 0.1, 1, 5 and 10) and fed perfect SAP-to-inhabitant
CSI to the optimal BF. We plot the average capacity of SSSC and the
average interference to inhabitants on each subcarrier which are obtained
by using 2000 Monte-Carlo realizations in Figure \ref{fig3:Average-capacity vs interference threshold}
and Figure \ref{fig4:Average-interference vs Inter threshold} respectively.

Figure \ref{fig3:Average-capacity vs interference threshold} reveals
the following information. 1. For the system adopts BD precoding including
inhabitants, the average capacity of SSSC shakes off the shackles
of interference threshold within the small radius of uncertainty region.
For $\epsilon^{2}=0.1$, the average capacity of SSSC achieves 54.7651
bps when the interference threshold is 1.75W. After that it oscillates
on small scale and no longer increases with the enhancement of threshold.
This is because BD precoding based on uncertain CSI can inhibit the
interference below the interference threshold within the small radius
of uncertainty region. Figure \ref{fig4:Average-interference vs Inter threshold}
exhibits that the average interference to inhabitants on each subcarrier
for $\epsilon^{2}=0.1$ is maintained below the threshold when the
threshold is no less than 0.25W. For $\epsilon^{2}=0$, since BD precoding
based on perfect CSI, the interference is completely eliminated and
accordingly the average capacity of SSSC has no relationship with
the interference threshold. It fluctuates slightly with the fluctuation
of channel between SAP and sojourners. For $\epsilon^{2}=1,\;5,\;\mathrm{and}\;10$,
the auxiliary optimal BF plays the leading role in inhibiting the
interference. The actual interference is always equal to the threshold.
The average capacity of SSSC exhibits uptrend as the threshold is
enhanced. 2. System adopts BD precoding including inhabitants achieves
higher capacity compared with that exploits BD precoding excluding
inhabitants within a considerably large radius of uncertainty region.
The average capacity of SSSC of the first system declines as the radius
of uncertainty region increases. It should be noted that until $\epsilon^{2}$
increases to $10$, the average capacity of SSSC of the first system
is still higher than that of the second system with $\epsilon^{2}=0$.
This verifies Theorem 2.

\section{Conclusion}

In regard to design issues in CSCS, we put forward practical propositions
for determining the number of transmit antennas and controlling downlink
inter-sub-small cell interference. We found how the factors influence
the number of concurrent sojourners and further impact the number
of transmit antennas on sub-small cell APs. We proved and verified
that \textit{BDBF} performs better than the interference controller
using optimal BF alone for gaining more capacity within a bearably
large radius of uncertainty region.  

\section*{Acknowledgements} This research is supported by SACRA project (FP7-ICT-2007-1.1, European Commission-249060).

\end{document}